\documentclass[
 aps,
 amsmath,amssymb,superscriptaddress,
 reprint,
]{revtex4-1}

\usepackage{graphicx}% Include figure files
\usepackage{hyperref,amsmath}
\usepackage{dcolumn}% Align table columns on decimal point
\usepackage{bm}% bold math
\begin{document}

\title[]{Circuit-quantum electrodynamics with direct magnetic coupling to single-atom spin qubits in isotopically enriched $^{28}$Si}

\author{Guilherme Tosi}
 \email{g.tosi@unsw.edu.au}
 \affiliation{Centre for Quantum Computation and Communication Technology, School of Electrical Engineering \& Telecommunications, UNSW Australia, Sydney, New South Wales 2052, Australia.}
\author{Fahd A. Mohiyaddin}
 \affiliation{Centre for Quantum Computation and Communication Technology, School of Electrical Engineering \& Telecommunications, UNSW Australia, Sydney, New South Wales 2052, Australia.}
\author{Hans Huebl}
 \affiliation{Walther-Mei$\beta$ner-Institut, Bayerische Akademie der Wissenschaften, D-85748 Garching, Germany.}
 \affiliation{Nanosystems Initiative Munich (NIM), Schellingstr. 4, D-80799 Munich, Germany.}
\author{Andrea Morello}
 \email{a.morello@unsw.edu.au}
 \affiliation{Centre for Quantum Computation and Communication Technology, School of Electrical Engineering \& Telecommunications, UNSW Australia, Sydney, New South Wales 2052, Australia.}

\date{\today}

\begin{abstract}

Recent advances in silicon nanofabrication have allowed the manipulation of spin qubits that are extremely isolated from noise sources, being therefore the semiconductor equivalent of single atoms in vacuum. We investigate the possibility of directly coupling an electron spin qubit to a superconducting resonator magnetic vacuum field. By using resonators modified to increase the vacuum magnetic field at the qubit location, and isotopically purified $^{28}$Si substrates, it is possible to achieve coupling rates faster than the single spin dephasing. This opens up new avenues for circuit-quantum electrodynamics with spins, and provides a pathway for dispersive read-out of spin qubits via superconducting resonators.

\end{abstract}

\maketitle

\section{Introduction}

Natural atoms in vacuum are the cleanest and most reproducible quantum systems, but they pose limitations to the way they can be made to interact with their environment. In the cavity-Quantum Electrodynamics (cavity-QED) scheme, atoms interact with photons in a high-finesse cavity, but the strength and duration of the interaction is limited by the electric dipole of the atoms and the dwell time in the cavity \cite{Walther2006}. Ten years ago, the progress in nanofabrication and in coherent control of nanoscale electrical circuits opened a new avenue in this field, known as circuit-QED \cite{Schoelkopf2008,You2011,Xiang2013}. Large artificial atoms are fabricated with superconducting thin films and Josephson junctions, and coupled to the quantized electromagnetic modes of a high-Q on-chip superconducting resonator. The dipole moment can be made almost arbitrarily large, and the dwell time is infinite. Unlike natural atoms, it is very easy to tune in-situ the energy spectrum and various other properties of artificial atoms. This architecture has brought about some of the most exquisite demonstrations of control over individual and multiple quantum systems, including quantum logic gates \cite{DiCarlo2009} and quantum teleportation \cite{Steffen2013}. Because of their large size, and the presence of amorphous materials and interfaces in their vicinity, the superconducting qubits used in circuit-QED are not the most long-lived quantum systems. Their lifetime has steadily improved over the years, reaching up to $10$~$\mu$s in 3D cavities \cite{Paik2011}, but still does not match that of true atomic systems. The ``ultimate setup'' in this field would be to combine the purity of atoms in vacuum with the convenience
and tuneability of circuits in solids.

The term ``semiconductor vacuum'' \cite{Steger2012} has been adopted to describe the exceptional properties of isotopically purified $^{28}$Si. Ultra-high purity samples are being produced for the purpose of redefining the kilogram in the SI units \cite{Becker2012}, but they are also used as hosts for the most coherent quantum systems demonstrated so far in solid state. A substitutional group V donor atom in Si (such as P, As, Sb or Bi) behaves to a good approximation like hydrogen in vacuum, with an energy spectrum renormalized by the effective mass and dielectric constant of the host material \cite{Greenland2010}. The absence of nuclear spins and paramagnetic states in $^{28}$Si implies that the electron and nuclear spins of a donor atom behave almost as if they really were held in a magnetic vacuum. Indeed, extraordinary coherence times have been measured in bulk samples for both the electron ($T_{2e}=10$~s \cite{Tyryshkin2012}) and the nucleus ($T_{2n}=3$ hours \cite{Saeedi2013}) of $^{31}$P donor atoms in $^{28}$Si.  Moreover, the weakness of spin-orbit coupling in P donors \cite{Mayur1993} makes the donor electron insensitive to electric field fluctuations, tremendously reducing the impact of charge noise so common in nanostructures.

In this paper we investigate the possibility of using the spin of a $^{31}$P donor atom in $^{28}$Si to realize the ultimate circuit-QED setup -- coupling a single atom in solid state to a single photon in a microwave circuit. In contrast to recent proposals in which the electron is coupled to the resonator electric field via different spin-orbit interaction mechanisms \cite{Burkard2006,Abanto2010,Jin2012,Hu2012,Cottet2010}, here we consider the case where the coupling is directly provided via the resonator magnetic field. 

\newpage
\section{Spin-resonator coupling}

The interaction between an electron spin-1/2 and a photonic mode confined inside a resonator is described by the Jaynes-Cummings Hamiltonian:

\begin{equation}
H=\frac{\epsilon_z}{2}\sigma_z+h\nu_0\left(a^\dagger a+\frac{1}{2}\right)+g\sigma_x(a^\dagger+a),
\end{equation}

where $\epsilon_z$ is the electron Zeeman energy, $\nu_0$ the photon frequency and $g$ the coupling constant.

We consider an electron bound to a $^{31}$P dopant under an applied constant magnetic field $B_0$. In this case the electron Zeeman energy is $\epsilon_z=h\gamma_eB_0+A/2\cdot\sigma_z^n$, with $\gamma_e=28$~GHz/T being the electron gyromagnetic ratio, $A$ the electron nucleus hyperfine coupling and $\sigma_z^n$ the Pauli operator for the nucleus spin state. In this paper we consider $B_0>100$~mT in such a way that the hyperfine coupling, on the order of $0.1$~GHz, is much smaller than the electron Zeeman splitting. This removes any entanglement between the electron and the nuclear spin states. Moreover, since the nuclear spin state lifetime is in the order of many hours \cite{Muhonen2014}, $\epsilon_z$ can be assumed as a constant in the Hamiltonian, and the nuclear degree of freedom neglected.

The spin-photon coupling rate $g/h$ is assumed to be equal to half the Rabi frequency of an electron under the resonator magnetic vacuum field, which has amplitude $B_{\rm vac}$ and direction perpendicular to $B_0$:

\begin{equation} \label{eq:gBvac}
\frac{g}{h}=\frac{\gamma_eB_{\rm vac}}{4}
\end{equation}

\begin{figure}
\centering
\includegraphics[width=0.48\textwidth]{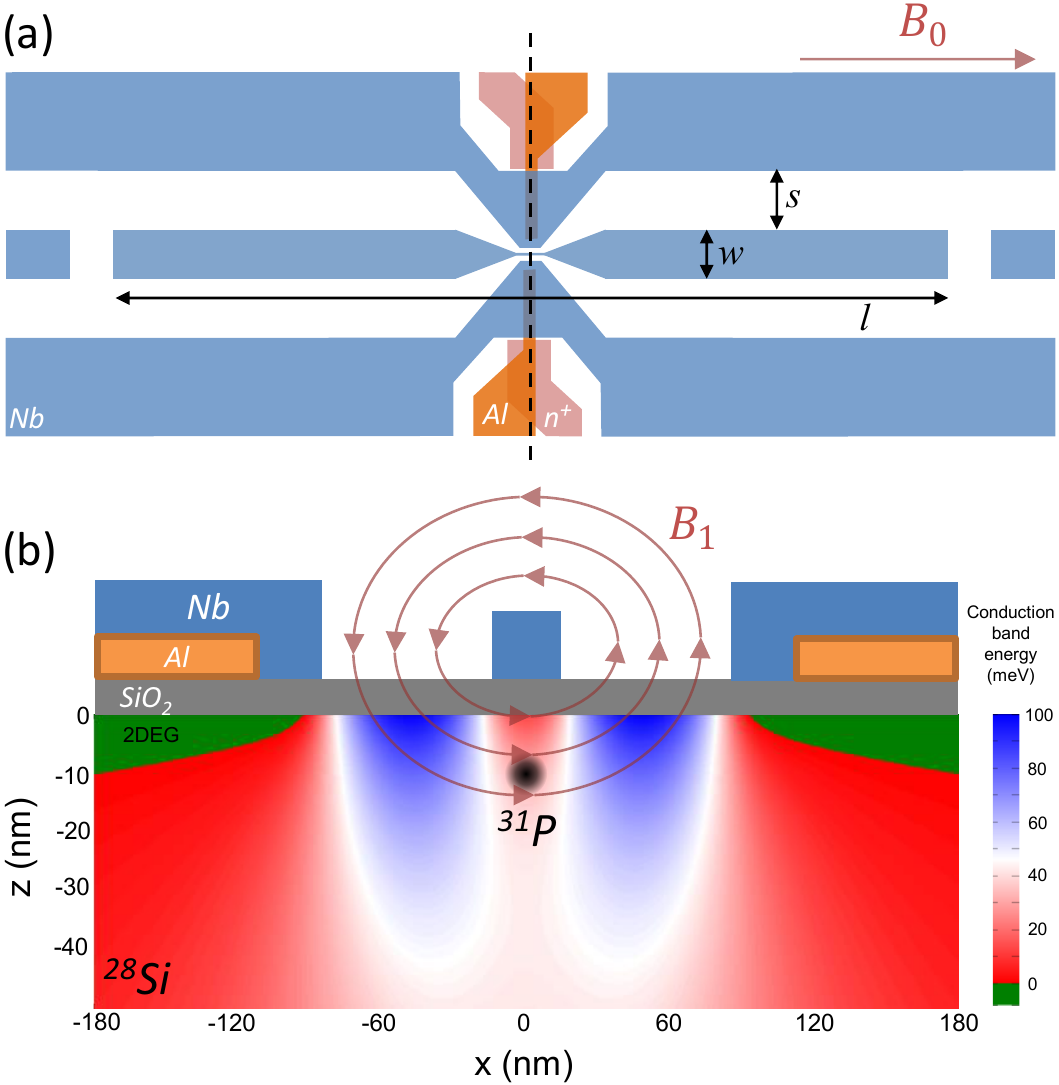}
\caption{Device architecture. (a) Coplanar waveguide resonator, with a constriction at its center. The $n^+$ region below the SiO$_2$ layer and the Al top gate are schematically shown, as well as the direction of the magnetic field $B_0$, applied in-plane to minimize the creation of quantized flux lines in the superconducting Nb film. (b) TCAD simulation of the conduction band profile $E_c$, for a cross section of the device depicted by a dashed black line in (a). Values are plotted with respect to the Fermi level $E_F$ of the $n^+$ reservoir. The green regions, where $E_c < E_F$, show extension of the induced electron layer. The SiO$_2$ layer, Al top gates, and Nb ground plane and center lines, and resonator magnetic field $B_1$ are also schematically shown. Voltages used are $V_{\rm Al}=+1.63$~V, $V_{\rm Nb}=0$~V and $V_{n^+}=-0.37$~V. Material thicknesses are $t_{\rm SiO_2}=8$~nm, $t_{\rm Nb}=20$~nm, $t_{\rm Al}=16$~nm and $t_{\rm Al_2O_3}=2$~nm. Widths are $w=30$~nm and $s=70$~nm. The Nb ground plane extends inwards 15~nm beyond the Al gate. In the TCAD model We assume a Si/SiO$_2$ interface charge density of $-2\times10^{11}$~cm$^{-2}$, consistent with previous estimates in similar devices \cite{Mohiyaddin2013,Yang2013,McCallum2008}. The temperature assumed in the TCAD model is 1~K (numerical convergence becomes problematic below this value). Note that the $z$ and the $x$ axis have a different scale. The location of the $^{31}$P donor is shown, as well as the straylines of the waveguide magnetic field $B_1$. In the absence of drive, $B_1=B_{\rm vac}$.}
\label{fig:device}
\end{figure}

In order to calculate $g$, we therefore have to calculate the strength of the resonator magnetic vacuum field. We consider a resonator in which the central line of the coplanar waveguide (CPW) is capacitively interrupted at two points separated by a distance $l$ (Fig. \ref{fig:device}a). The resonator transmits signals whose frequencies are integer multiples of the fundamental mode, which is the one whose half-wavelength is equal to the resonator length, $\lambda_0/2=l$. This condition implies the frequency of the fundamental mode to be:

\begin{equation} \label{eq:nu0}
\nu_0=\frac{c}{\sqrt{\epsilon_{\rm eff}}}\frac{1}{2l},
\end{equation}

where $\epsilon_{\rm eff}$ is the effective dielectric constant of the CPW and $c$ the speed of light. The magnetic field profile of this mode is maximum in the center of the resonator, which is where the $^{31}$P donor has to be placed (Fig. \ref{fig:device}a). 

To avoid losses, the waveguide layer is made of a superconducting material, e.g. Nb. This layer sits on a few-nanometers-thick layer of SiO$_2$ followed by a $^{28}$Si substrate. For the present purpose, it is perfectly acceptable to use an isotopically enriched epilayer of $\sim 1$~$\mu$m thickness, grown on top of natural silicon \cite{Muhonen2014}. In this limit where the substrate is much thicker than any other layer, $\epsilon_{\rm eff}\approx \frac{1+\epsilon_s}{2}=6.3$ \cite{Gevorgian1995}, where $\epsilon_s=11.6$ is the silicon dielectric constant.

The value of $B_{\rm vac}$ depends on the amplitude of the zero-point current in the resonator, $I_{\rm vac}$. The latter can be calculated by assuming the energy of the vacuum field to be stored in the resonator equivalent lumped inductance $L$:

\begin{equation}\label{eq:VacEn}
\frac{h\nu_0}{2}=\frac{L{I_{\rm vac}}^2}{2}
\end{equation}

The equivalent inductance of the resonator, for the fundamental mode, is known to be \cite{Pozar2005,Goppl2008}:

\begin{equation}\label{eq:EqInd}
L=\frac{Z_0}{\pi^2\nu_0},
\end{equation}

where $Z_0$ is the characteristic impedance of the transmission line, which depends on the line width $w$ and gap $s$ (Fig. \ref{fig:device}a). Impedance matching to the outside circuitry requires $Z_0=50~\Omega$. From Eqs. \ref{eq:VacEn} and \ref{eq:EqInd}, we find the resonator vacuum current:

\begin{equation}\label{eq:Ivac}
I_{\rm vac}=\pi\sqrt{\frac{h}{Z_0}}\nu_0\Rightarrow I_{\rm vac}[{\rm A}]\approx1.14\times10^{-17}\nu_0[{\rm Hz}]
\end{equation}

For conventional CPW resonators used in circuit-QED experiments, the central line is wide enough so that one can consider the vacuum magnetic field a few nanometers underneath to be proportional to the vacuum current density $I_{\rm vac}/w$ \cite{Griffiths1999},

\begin{equation}\label{eq:Bvac_plane}
B_{\rm vac}\approx\frac{\mu_0}{2}\frac{I_{\rm vac}}{w},
\end{equation}

where $\mu_0$ is the vacuum permeability. For $w=20~\mu$m and $B_0=200$~mT ($\nu_0=5.6$~GHz), $B_{\rm vac}\approx2$~nT, yielding spin-photon coupling rates $g/h=\gamma_eB_{\rm vac}/4\approx14$~Hz.

In order to increase the resonator vacuum field at the donor location, therefore increasing the coupling strength, we propose to shrink the central line width as to increase the vacuum current. A similar procedure has been used to couple resonator magnetic fields to flux qubits \cite{Abdumalikov2008}, achieving coupling rates as high as to reach the ultrastrong coupling regime \cite{Niemczyk2010}. Here we assume $w=30$~nm, compatible electron beam lithography techniques. In order to avoid losses, the characteristic impedance of the constricted region has to be kept $Z_0=50~\Omega$ which implies a transmission line gap width $s=70$~nm \cite{Simons2001}. Such a small gap, if constant along all the resonator length, would result in very high electric fields $E_{\rm vac}=V_{\rm vac}/s$, which can greatly deteriorate the resonator Q-factor by driving dissipative dynamics of charge fluctuators in and around the gap. Note however that our region of interest is only at the center of the resonator length, where the magnetic vacuum field has an antinode and the electric field a node. In this region, electric losses are negligible. We therefore choose the $s=70$~nm gap to be localized in a constriction where the donor is to be implanted, and $s=10$~$\mu$m everywhere else (Fig. \ref{fig:device}a).

Considering Nb film thickness $t_{\rm Nb}=20$~nm and a donor implanted $20$~nm below the oxide-Nb interface, $B_{\rm vac}$ at the donor location is approximately given by the Ampere's law,

\begin{equation}\label{eq:Bvac_wire}
B_{\rm vac}\approx\frac{\mu_0I_{\rm vac}}{2\pi r},
\end{equation}

where $r\approx30$~nm is the distance between the donor and the center of the central line (Fig. \ref{fig:device}b). Since the zero-point current is given by Eq.~\ref{eq:Ivac}, the spin-resonator coupling, $g/h=\gamma_eB_{\rm vac}/4$, is also found to depend linearly on the frequency,

\begin{equation}\label{eq:g}
g/h\approx 5.3\times10^{-7}\cdot\nu_0
\end{equation}

In order to achieve the strong coupling regime, $g$ has to be higher than the qubit dephasing, $\gamma^*$, and photon decay, $\kappa=\frac{\nu_0}{Q}$, where $Q$ is the resonator Q-factor. The inhomogeneous linewidth of $^{31}$P single electron spins in isotopically purified $^{28}$Si has been recently measured to be as low as $\gamma^*=1.2$~kHz, corresponding to a Ramsey dephasing time $T_2^*=270~\mu$s \cite{Muhonen2014} (to be compared e.g. to $T_2^*=5$~ns measured in GaAs dots \cite{Petta2005}). Furthermore, if we assume $B_0=200$~mT ($\nu_0 = 5.6$~GHz resonator frequency as to have spin-photon resonance), one could get $g \approx 3$~kHz~$> \gamma^*$. The corresponding vacuum current is $I_{\rm vac} \approx 64$~nA, or a current density $j_{\rm vac}=I_{\rm vac}/(20\times30\,\mathrm{nm}^2)\approx 10^4$~A/cm$^2$. This is two orders of magnitude less than the known critical current density in Nb thin films \cite{Huebener1975,Kim2009}, ensuring there is no risk of breaking the superconducting state of the central line. 

The other requirement, $g > \kappa$, translates into $Q > 2 \times 10^6$. Even though such high-Q resonators are feasible \cite{Day2003,Megrant2012}, the presence of magnetic fields ($B_0 \approx 200$~mT) is likely to introduce extra losses through the creation of vortices in the superconducting film. We find therefore the peculiar situation where it is the cavity decay $\kappa$ instead of the qubit dephasing $\gamma^*$ that poses the greatest hurdle to achieving the strong coupling regime.

In addition to the microwave engineering aspects, this architecture also requires ensuring that there is one and only one electron bound to the $^{31}$P donor. For a donor near (e.g. $\simeq 20$~nm under) a Si/SiO$_2$ interface, fixed charge in the SiO$_2$ and at the Si/SiO$_2$ interface above the donor can lift its electrochemical potential $\mu_D$ and lead to donor ionization \cite{Rahman2009}. To circumvent this problem we consider the addition of an electron reservoir in the vicinity of the donor. The reservoir is induced with the help of an aluminum `top-gate', held at voltage $V_{\rm top}$  (beneath the Nb ground plane in Fig. \ref{fig:device}b), which attracts electrons from a heavily doped $n^+$ source region (Fig. \ref{fig:device}a), held at voltage $V_s$. The electron reservoir is induced when $V_{\rm top} - V_s$ is larger than some threshold (typically around 0.6~V), but both voltages can float with respect to ground. Here, $V = 0$ ground is the potential of the resonator ground planes and center conductor. Therefore, it is possible to choose $V_s$ such that the reservoir Fermi level $E_F$ is higher than $\mu_D$, and ensure that the donor is neutral. We note that the donor-reservoir distance $\approx 70$~nm is larger than the typical distances $\approx 25$~nm used in donor-qubit devices \cite{Mohiyaddin2013}. However this is not an issue, because the reservoir's only role here is to ensure donor charge neutrality -- we do not seek to produce fast spin-dependent tunneling events between donor and reservoir to achieve spin readout \cite{Morello2010}.

In Figure \ref{fig:device}b, we plot the conduction band energy $E_c$ -- computed with TCAD \footnote{{http://www.synopsys.com/TOOLS/TCAD/}} --  along a slice of the device, having set the reservoir Fermi level $E_F$ as the zero-energy reference. We set the Nb ground planes and center conductor at ground ($V=0$) and choose $V_{\rm top}=+1.63$~V, with $V_s = -0.37$~V. The electrochemical potential $\mu_D^0$ of the neutral ($D^0$) donor charge state is $\approx 45.6$~meV below the conduction band edge $E_c$ \cite{Kohn1955}, assuming negligible Stark shifts in our nanostructure \cite{Mohiyaddin2013,Rahman2009}. A second electron can be added to the donor creating the negatively-charged $D^-$ state, at the electrochemical potential $\mu_D^- \approx E_c - 10$~meV \cite{Tan2010} (this value can vary by several meV depending on the electrostatic environment of the donor). Therefore, the donor is expected to be in the neutral $D^0$ charge state whenever the conduction band edge at the donor location is between $\sim 10$ and $45.6$~meV above the reservoir $E_F$. As shown in Fig. \ref{fig:device}b, this condition is satisfied for a wide range of donor locations, including the region underneath the resonator central line, where $B_1$ is maximum. 

\section{Spin control and read-out}

In order to avoid spin-to-photon conversion while performing quantum gate operations on the electron spin, it is convenient to detune the spin Larmor frequency from the resonator mode. We therefore assume $\Delta\gg g$, where $\Delta=h\nu_0-\epsilon_z$.

In this so-called dispersive regime, the diagonalized Hamiltonian is approximately \cite{Blais2004}:

\begin{equation}\label{eq:DispHam}
H\approx\frac{\epsilon_z}{2}\sigma_z+\left(h\nu_0+\frac{g^2}{\Delta}\sigma_z\right)\left(a^\dagger a+\frac{1}{2}\right)
\end{equation}

The corresponding eigenstates are approximately the same as the uncoupled Hamiltonian, with a small deviation proportional to $(g/\Delta)^2$ \cite{Haroche1992}. In order to have eigenstates with $99\%$ fraction of uncoupled modes, therefore protecting the qubits from decaying into photons, we will assume from now on:

\begin{equation}\label{eq:D10g}
\Delta=10g
\end{equation}

Importantly, Eq. \ref{eq:DispHam} implies that the cavity resonance depends on the spin state. Therefore, the measurement of the cavity transmission with a weak microwave signal allows for the quantum non-demolition readout of the spin state. Conversely the only spin readout method demonstrated so far with donor spins \cite{Morello2009,Morello2010} causes the physical loss of the electron upon readout.

The spin-dependent cavity resonance shift can be measured through the resonator phase-shift, whose spin-state dependent values are $\pm \arctan{\left(\frac{2g^2Q}{h\nu_0\Delta}\right)}$ \cite{Wallraff2004}. Assuming phase sensitivity on the order of $0.1^{\circ}$ for homodyne-detection setups \cite{Petersson2012} and Eq. \ref{eq:D10g}, the qubit state can be measured for Q-factors as low as $Q=1.6\times10^4$, which is consistent with routine CPW resonators.

Note that we have not assumed any operation frequency when deriving the limit for the Q-factor. Indeed, Eqs. \ref{eq:g} and \ref{eq:D10g} impose that the spin-dependent cavity shift is $\pm \arctan{\left(1.07\times10^{-7}\cdot Q\right)}$. This means that it is possible to choose a relatively low operation frequency $\nu_0$, corresponding to a small static field $B_0$ and therefore optimal Q-factor, improving the qubit read-out fidelity. 

\begin{figure*}
\centering
\includegraphics[width=0.8\textwidth]{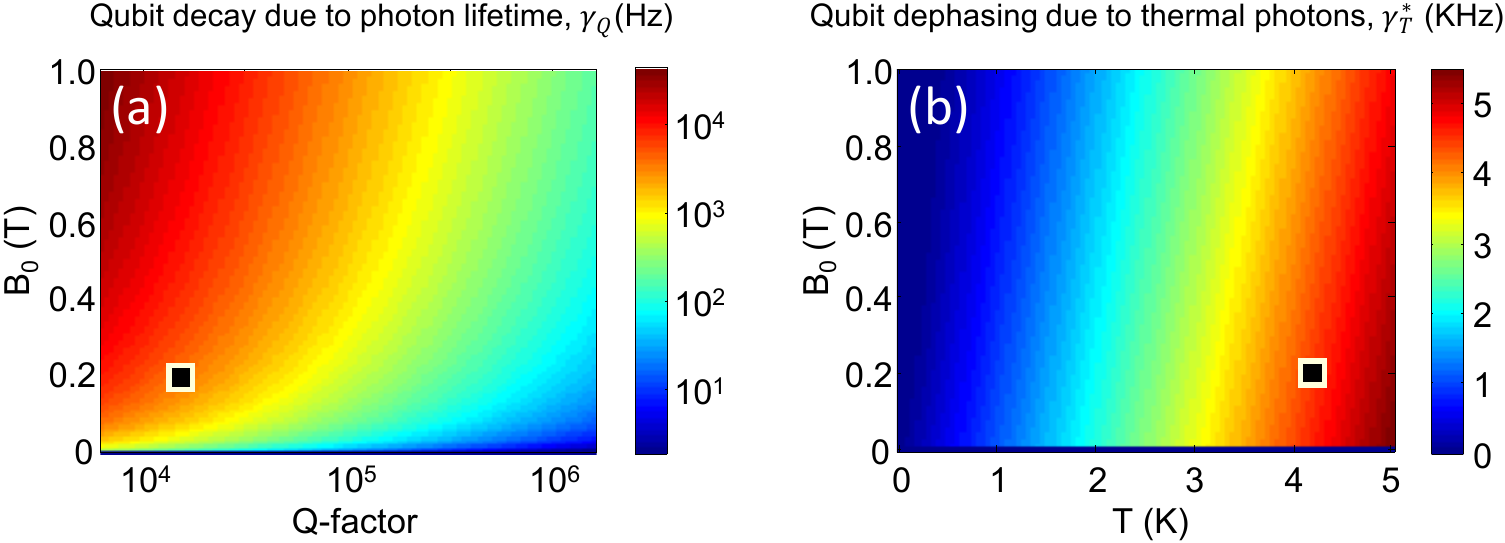}%,clip=true,trim=0cm 13.1cm 8.6cm 0cm
\caption{Enhancement of the electron spin dephasing rate as predicted by (a) Eq. \ref{eq:gammaQ} and (b) Eq. \ref{eq:gammaT}. Parameters for a realistic device are depicted by black squares: (a) $\gamma_Q(Q=1.5\times10^4,B_0=200~{\rm mT})=3.5~{\rm kHz}$ and (b) $\gamma_T^*(T=4.2~{\rm K},B_0=200~{\rm mT})=4.5~{\rm kHz}$.}
\label{fig:dephasing}
\end{figure*}

The operation frequency is also important in determining the enhancement of the spin decay rate due to its coupling to resonator photons with finite lifetime. This enhancement is simply given by the photon fraction of the Hamiltonian (Eq. \ref{eq:DispHam}) eigenstates times the photon decay rate \cite{Blais2004}:

\begin{equation}\label{eq:gammaQ}
\gamma_Q=\left(\frac{g}{\Delta}\right)^2\frac{\nu_0}{Q},
\end{equation}

which is equal to $\nu_0/(100Q)$ for our choice of detuning $\Delta = 10g$ (Eq. \ref{eq:D10g}). Such a dependence is plotted in Fig. \ref{fig:dephasing}a. For instance, at an operating frequency $\nu_0=5.6$~GHz (corresponding to $B_0=200$~mT), a quality factor as low as $Q=1.5\times10^4$ yields an increase in spin decay equal to $3.5$~kHz (black square in Fig. \ref{fig:dephasing}a), which is of the same order as the intrinsic dephasing $\gamma^*$ of the isolated electron spin.

Another source of dephasing comes from thermal fluctuations of the photon number in the resonator. Indeed, the terms in Eq. \ref{eq:DispHam} can be rearranged as to highlight that the spin resonance depends on the photon number, $\epsilon_z\rightarrow\epsilon_z+\left(g^2/\Delta\right)(2a^\dagger a+1)$. This implies that the spin resonance linewidth, and therefore the qubit dephasing rate, increases with thermal photon occupation. The photon number in the fundamental mode \footnote{we neglect higher resonator modes since the qubit is hugely detuned from those} is given by the Bose-Einstein distribution, $n=(e^{h\nu_0/k_BT}-1)^{-1}$. The spin dephasing is therefore increased by $\gamma_T^*=\left(g^2/\Delta\right) n$, which can be written, recalling Eqs. \ref{eq:g} and \ref{eq:D10g}, as:

\begin{equation}\label{eq:gammaT}
\gamma_T^*=\frac{5.3\times10^{-8}\cdot\nu_0}{e^{h\nu_0/k_BT}-1}
\end{equation}

Such a dependence is plotted in Fig. \ref{fig:dephasing}b for a range of temperatures and operating frequencies. The enhanced spin dephasing remains on the order of its uncoupled dephasing rate for temperatures up to liquid helium (4.2 K), for all ranges of operating frequencies.

Note that the spin-dependent cavity shift does not depend on the photon number inside the resonator, and therefore the effectiveness of the readout method should not depend on temperature (until the superconducting resonator starts to degrade). Moreover, the relaxation rate $T_1^{-1}$ of the electron spin at high temperatures does not limit its dephasing, since it remains slower than 1 Hz up to $T \approx 4$~K \cite{Tyryshkin2012}. As mentioned before, $\nu_0$ and thus $B_0$ can be kept low, further decreasing the spin relaxation rate $T_1^{-1} \propto B^5$ \cite{Morello2010}.

The electron spin state can be rotated by applying to the resonator a microwave pulse with the same frequency as the AC Stark-shifted spin Larmor frequency. Note that high input powers have to be used, since the drive is out-of-resonance with the resonator and therefore it is mainly reflected at the input port. The maximum Rabi-frequency of the electron spin is given by the critical current density in the center line before superconductivity is lost. The critical current of niobium films, $1-10\times10^{6}$~A/cm$^2$ \cite{Huebener1975,Kim2009}, is enough to drive the spin at $1-10$~MHz rates, there orders of magnitude faster than its dephasing rate.

\section{Cavity-mediated macroscopic entanglement}

\begin{figure*}
\centering
\includegraphics[width=0.8\textwidth]{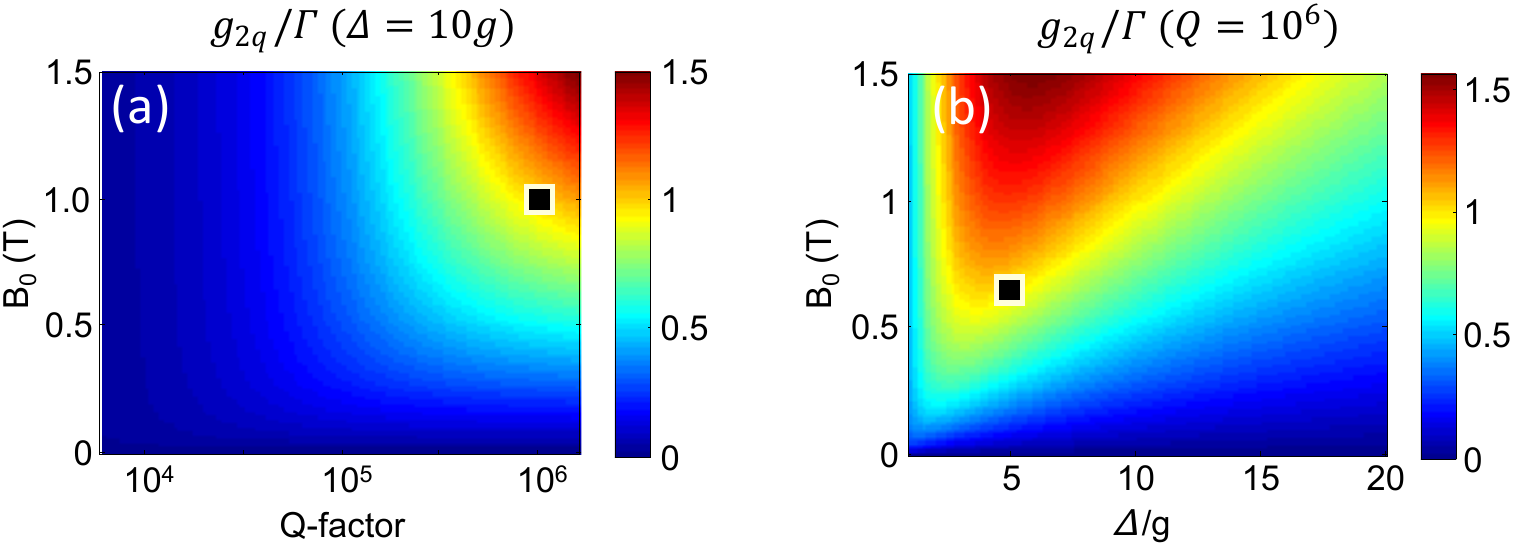}%,clip=true,trim=0cm 13.5cm 10.1cm 0cm
\caption{Ratio between coupling strength between qubits and their linewidth, $\frac{g_{2q}}{\Gamma}$, as predicted by Eq. \ref{eq:g2qGamma}. In (a), the spin-photon detuning is fixed at $\Delta=10g$ whereas the Q-factor and the magnetic field $B_0$ (proportional to the operating frequency $\nu_0$) vary. In (b), the resonator Q-factor is fixed at $Q=10^6$ whereas the detuning $\Delta$ and the magnetic field $B_0$ vary. Parameters depicted by black squares are: (a) $\frac{g_{2q}}{\Gamma}(Q=10^6,B_0=1~{\rm T},\Delta=10g)=1$ and (b) $\frac{g_{2q}}{\Gamma}(Q=10^6,B_0=650~{\rm mT},\Delta=5g)=1$.}
\label{fig:swap}
\end{figure*}

One of the greatest advantages of coupling qubits to CPW resonators is that the latter can be used as a bus to entangle qubits placed at different points along the resonator, and therefore separated by macroscopic distances. This is also done in the dispersive regime, with the coupling provided by virtual photons \cite{Majer2007}. The qubit-qubit coupling strength is then given by \cite{Blais2004}:

\begin{equation}\label{eq:gamma2q}
g_{2q}=g^2/\Delta
\end{equation}

For our chosen set of parameters (Eqs. \ref{eq:g} and \ref{eq:D10g}), this coupling is proportional to the resonator frequency, $g_{2q}/h\approx5.3\times10^{-8}\nu_0$. In order to have a macroscopic coupling rate higher than the intrinsic qubit dephasing rate, $g_{2q}>\gamma^*$, operating frequencies $\nu_0>22.5$~GHz ($B_0>800$~mT) are therefore required. Note however that such high frequencies would also increase the qubit decay rate induced by photon losses (see Eq. \ref{eq:gammaQ} and Fig. \ref{fig:dephasing}). One therefore has to carefully choose the set of parameters that maximizes the ratio $g_{2q}/\Gamma$, where $\Gamma$ is the total single qubit linewidth. Let us neglect the qubit dephasing due to thermal photons, $\gamma_T^*$ (Eq. \ref{eq:gammaT}), by noting that it does not depend much on $B_0$ (Fig. \ref{fig:dephasing}) and that it is negligible for temperatures below a few hundred millikelvin. We therefore have $\Gamma=h(\gamma^*+\gamma_Q)$. Finally, we will consider $\Delta$ as a optimization parameter, since $g_{2q}$ and $\gamma_Q$ depend differently on it (linearly for $g_{2q}$, Eq. \ref{eq:gamma2q}, and quadratically for $\gamma_Q$, Eq. \ref{eq:gammaQ}). The coupling to linewidth ratio therefore can be written as:

\begin{align}\label{eq:g2qGamma}
\frac{g_{2q}}{\Gamma}=\frac{g_{2q}}{h(\gamma^*+\gamma_Q)}=\frac{g^2}{h\Delta\left[\gamma^*+\left(\frac{g}{\Delta}\right)^2\frac{\nu_0}{Q}\right]} \nonumber \\
=\frac{h(5.3\times10^{-7}\cdot\nu_0)^2}{\Delta\left[1.2~\mathrm{kHz}+\left(\frac{5.3\times10^{-7}\cdot\nu_0}{\Delta}\right)^2\frac{\nu_0}{Q}\right]}
\end{align}
%\begin{equation}\label{eq:g2qGamma}
%\frac{g_{2q}}{\Gamma}=\frac{g_{2q}}{h(\gamma^*+\gamma_Q)}=\frac{g^2}{h\Delta\left[\gamma^*+\left(\frac{g}{\Delta}\right)^2\frac{\nu_0}{Q}\right]}
%=\frac{h(5.3\times10^{-7}\cdot\nu_0)^2}{\Delta\left[1.2~\mathrm{kHz}+\left(\frac{5.3\times10^{-7}\cdot\nu_0}{\Delta}\right)^2\frac{\nu_0}{Q}\right]}
%\end{equation}

We first set the qubit-photon detuning to our previous assumption, $\Delta=10g$, and plot the $g_{2q}/\Gamma$ dependence on $\nu_0$ and $Q$ in Fig. \ref{fig:swap}a. As expected, the ratio increases with magnetic field and Q-factor, being equal to one for $Q=10^6$ and $B_0=1$~T. Even though such a high field is below the critical one that breaks up superconductivity of Nb films \cite{Asada1969}, the high losses introduced by proliferation of vortices are likely to lower the resonator Q-factor by a significant amount. In the following we attempt to lower the need for high $B_0$ by investigating the $g_{2q}/\Gamma$ dependence on spin-photon detuning $\Delta$. We assume $Q=10^6$ and plot, in Fig. \ref{fig:swap}b, the dependence of Eq. \ref{eq:g2qGamma} on $\nu_0$ and $\Delta$. For a fixed $B_0$-field, we see that $g_{2q}/\Gamma$ increases with $\Delta$, which is expected since $g_{2q}$ decreases linearly with $\Delta$ whereas $\Gamma$ decreases quadratically. After a maximum detuning, however, the ratio $g_{2q}/\Gamma$ starts decreasing again. This happens whenever $\gamma_Q<\gamma^*$ and therefore the intrinsic spin dephasing rate $\gamma^*$ is the main loss channel. At this point, $g_{2q}$ decreases with $\Delta$ whereas $\Gamma$ is unaffected. We find an optimal operating point at $\Delta=5g$ and $B_0=650$~mT (black square), at which $g_{2q}/\Gamma=1$. We note however that operating at such small detuning decreases the entanglement fidelity, since the spin eigenstantes of the Hamiltonian in Eq. \ref{eq:DispHam} contain 4\% ($g^2/\Delta=0.04$) of photon fraction.

\section{Conclusions and perspectives}

The architecture presented here takes full advantage of the exquisite isolation from the environment of a single electron spin bound to donor atoms in isotopically purified $^{28}$Si. Even though reaching the strong-coupling regime will be probably limited by the resonator Q-factor, coherent control and non-demolition readout of the qubit state can be performed via the resonator with no significant increase in the qubit dephasing, even for resonator Q-factors as low as $Q=10^4$ and liquid helium temperatures $T=4.2$~K.

The low spin-photon coupling rate makes however strong coupling of macroscopically separated qubits via virtual resonator photons extremely hard to achieve, also mainly due to expected low resonator Q-factors under high magnetic fields. A solution to this problem would be to introduce vortex pinning structures that limit their movement and therefore dissipation, increasing the resonator Q-factor \cite{Song2009,Bothner2011}. In this case it is desirable to have pinning centers whose size is on the order of the coherence length of Nb (around 40~nm \cite{Finnemore1966}) separated by a distance comparable to the London penetration depth (also around 40~nm \cite{Maxfield1965}). Therefore an array of nanoscale holes would be the optimum vortex trapping structure \cite{Latimer2013}. Note that here we propose to use Nb films whose thickness is smaller than the London penetration depth, and therefore cannot sustain a complete flux exclusion, resulting
in lower diamagnetic energy which then leads to a higher critical field \cite{Asada1969}. On the other hand the film thickness is also smaller than the Nb coherence length, which implies that the transition temperature will be slightly smaller \cite{Kim2009} .

Instead of relying on high Q-factors, one could look for resonator geometries that provide higher spin-photon coupling rates. An example is to introduce an artificial spin-orbit coupling \cite{Hu2012,Cottet2010} as to couple the spin state to the resonator electric field.

It is important to notice that the coupling rates derived in this paper rely on shrinking the resonator central line to a few tens of nanometers. Such a constriction, on the order of the Nb coherence length, will most certainly behave as a weak link and therefore determine a nanobridge-like Josephson junction \cite{Troeman2007,Mitchell2012}. Even though this increases the local inductance at the constriction \cite{Bourassa2009}, this is not associated with an increase of the magnetic vacuum field, since the junction inductance is purely kinetic and therefore not associated with any magnetic field. This is the reason why we ignored such an effect in this paper.

Finally, we note that the present proposal can apply also to electron spins in isotopically purified $^{12}$C, such as Nitrogen-Vacancy centers in diamond, which also can show intrinsic spin dephasing rates in the kHz range \cite{Balasubramanian2009}.

\section*{Acknowledgments}

We thank A. Laucht, J. T. Muhonen, J. P. Dehollain, R. Kalra and A. Blais for helpful discussions. This research was funded by the Australian Research Council Centre of Excellence for Quantum Computation and Communication Technology (project number CE110001027) and the US Army Research Office (W911NF-13-1-0024). H. H. acknowledges financial support by DFG (Grant No. SFB 631, C3).

\end{document}